\documentstyle[preprint,prl,aps,epsf]{revtex}
\begin{document}   
\draft

 \title{
Search for the $\pi$ Resonance in Two Particle 
Tunneling Experiments of $YBCO$ Superconductors}

\author{Y. Bazaliy, E. Demler, and S.C. Zhang}
 \address{
Department of Physics, Stanford University, Stanford, CA 94305 
} 
\date{\today}

\maketitle
 \begin{abstract}
 
A recent theory of the
resonant neutron scattering peaks in $YBCO$ superconductors
predicts the existence of a sharp 
spin triplet two particle collective mode (the ``$\pi$ resonance")
in the normal state. 
In this paper, we propose a experiment in which the $\pi$ resonance
could be probed directly in a two particle tunneling measurement.
 \end{abstract}

 \newpage

\narrowtext

Recent spin polarized inelastic neutron scattering experiments\cite{Fong,Dai,Rossat-Mignod}
of the $YBCO$ superconductor revealed the existence of a sharp 
collective mode. 
This collective mode has spin one, carries
momentum $(\pi,\pi)$ and has  well-defined energies
of $41meV$, $33meV$ and $25 meV$ respectively for 
materials with $T_c=92K$, $T_c=67K$ and $T_c=52K$ \cite{Fong}.
Most strikingly, this feature is
only observed in the neutron scattering experiment (of the
$T_c=92K$ and $T_c=67K$ materials) below the superconducting
transition temperature.

A number of theoretical explanations has been offered for this unique
feature\cite{neutron,other}. In particular, two of us \cite{neutron} 
proposed that there exists a sharp spin triplet particle-particle 
collective mode in
a wide class of strongly correlated models, including the Hubbard and
the $t-J$ model.  This collective mode, called the triplet $\pi$ mode
or simply the $\pi$ resonance\cite{so5}, is created by a two particle
operator
\begin{eqnarray}
\pi^\dagger= \sum_k (\cos k_x - \cos k_y) \;
      c_{k + Q,\uparrow}^\dagger 
      c_{-k,\uparrow}^\dagger
      \  \ Q=(\pi,\pi)
\label{pi}
\end{eqnarray}
It carries spin one, momentum
$(\pi,\pi)$ and its energy scale is determined by $J$, the spin
exchange energy. This collective mode
exists both in the normal and the superconducting state. In
the normal state, it does not couple to the neutron scattering probe
because of its particle-particle nature. However, below the superconducting
transition temperature $T_c$, this particle particle collective mode can mix
into the particle hole channel and couple to the neutron scattering
amplitude. Since the mixing amplitude is
proportional to the superconducting order parameter, this theory
offers a unique explanation why the resonant neutron scattering peak
disappears above the superconducting transition temperature. There is
a fundamental difference between this theory and the possible
alternative explanations based on the excitonic bound states inside
the superconducting gap.  In the former case, the existence of the
collective mode is not dependent on the superconducting order, only
the coupling to neutron is, while in the later case, the collective
mode will disappear entirely from the physical spectrum when the
superconducting gap disappears.

More recently, a unified theory of antiferromagnetism (AF) and $d$-wave
superconductivity (SC) in the high $T_c$ superconductors has been 
proposed\cite{so5}. This theory is based on a $SO(5)$ symmetry 
generated by the total spin, total charge and the $\pi$ operators
which rotate AF order parameters into the SC order parameters and
vice versa. Within this theory, the resonant neutron scattering peak
is interpreted as the pseudo-Goldstone boson associated with the
spontaneous breaking of the $SO(5)$ symmetry, reflecting the tendency
of a SC state to fluctuate into the AF direction. The $SO(5)$ symmetry
is based on the assumption that the $\pi$ operator is an approximate
eigenoperator of the microscopic Hamiltonian. Although there are
both analytical\cite{neutron} and numerical\cite{Meixner}
calculations in support of this assumption, it is certainly desirable
to test it in direct experiments.   

Therefore, in order to distinguish among the various theoretical
explanations of the resonant neutron peak, and to test the
$SO(5)$ theory of high $T_c$ superconductivity, it is crucial 
to search for the signature of the $\pi$
resonance in the normal state of the $YBCO$ superconductor. 
In the classic theoretical work of Scalapino\cite{Scalapino} and the
subsequent experimental confirmation by Goldman and 
co-workers\cite{Goldman}, 
a superconductor with a higher $T_c$ was used to probe the pairing
fluctuation of a lower $T_c$ material in the normal state. Inspired
by their ideas, we propose a
similar tunneling experiment to probe the $\pi$ resonance in both the
superconducting and the normal state. 
The proposed experimental geometry is depicted in Figure
1.  The proposed sample consists of a Josephson junction made out of a
lower $T_c$ superconductor (layer {\bf C}), a thin (less than the coherence
length) layer of a antiferromagnetic insulator (layer {\bf B}) and a bulk
higher $T_c$ material ({\bf A}), on the other side of
the junction. The lower $T_c$ and higher $T_c$ pairs of superconductors
can consist of a pair underdoped and optimally doped $YBCO$ superconductors
or a pair of optimally doped $YBCO$ and a $Ta$ or $Bi$ doped $BCO$
superconductors. The antiferromagnetic insulator can be realized by
the parent $YBCO$ insulator or the $Pr$ doped $BCO$ insulator.
All layers have their $ab$ plane perpendicular to the
tunneling direction. A voltage $V$ is applied in the tunneling
direction. 
\begin{figure}[t] 
\centerline{\epsfysize=4cm
\epsfbox{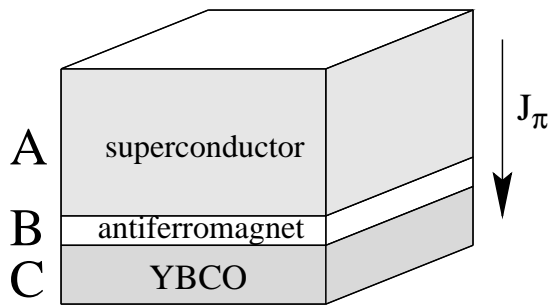}
}
\caption{Setting of the suggested experiment.}
\end{figure}
The basic idea is that when the temperature is in between
the two superconducting transition temperatures, the
BCS pairing condensate of the bulk superconductor {\bf A} acts as a
classical external field coupling to the two particle quantum operators
in the normal state of layer {\bf C}. 
The antiferromagnetic layer is
needed to transfer a center of mass momentum of $(\pi,\pi )$ to
the Cooper pair and flip its spin at the same time, so that it has
exactly the same quantum number as the $\pi$ resonance on the other
side of the junction.
 
To start with, let us consider the effective tunneling matrix element
between {\bf A} and {\bf C}, mediated through the antiferromagnetic
insulator. We can model the antiferromagnetic insulating layer {\bf B}
by a positive
$U$ Hubbard model at half-filling, with a one particle Greens function
given by\cite{swz}
\begin{eqnarray} 
G_{AF}(k,p;\alpha,\beta;\omega)
&=& \frac{(\omega+\epsilon_k)\delta_{\alpha\beta}\delta_{kp}
        + \Delta_{SDW} \delta_{\alpha, \; -\beta}\delta_{k,\; p+Q} }
       {\omega^2 - \Delta^2_{SDW} - \epsilon_k^2 } 
\label{Green}
\end{eqnarray}
where $\Delta_{SDW}$ is the spin-density-wave gap. We assume that
the chemical potentials of both {\bf A} and {\bf C} layers are within 
$\Delta_{SDW}$, the tunneling process is therefore non-resonant. From
(\ref{Green}) we see explicitly that electron can either tunnel
via a direct channel preserving it spin and transverse momentum, or
via a spin flip channel changing its transverse momentum by $Q$.
Consequently, we can model the tunneling from
{\bf A} to {\bf C} by the following effective Hamiltonian
\begin{eqnarray}
H_{T}=\sum_{pk\sigma} T^d_{pk}  a^{\dagger}_{p\sigma} c_{k\sigma} e^{i V t}
 + T^f_{pk} a^{\dagger}_{p+Q\sigma} c_{k-\sigma} e^{i V t} + h.c.
\label{explicit}
\end{eqnarray}
where $V$ is the applied voltage, the 
$a_{p\alpha}$ and $c_{k\alpha}$ operators refer
to the electronic operators in {\bf A} and {\bf C} with momenta
$p$ and $k$.
The ratio of the spin flip matrix element $T^f_{pk}$ to the
direct matrix element $T^d_{pk}$ is on the order of $\Delta_{SDW}/U$.
In the summation above $p$ is the usual three dimensional
momentum of electrons in bulk superconductor {\bf A}.  We model layer
{\bf C} as a two dimensional film with a 2-D momentum vector $k$.  

We consider a particular case of a perfectly specular scattering which
conserves the momentum parallel to the interface
\begin{eqnarray}
T^{d,f}_{kp} = T^{d,f} \delta_{k,p_{||}}
\label{spec}
\end{eqnarray}
The $\delta$ symbol above is a Kroneker delta of the 
discrete momenta, and $T^{d,f}$ are assumed to be constant.
In what follows we use common approximations in the theory  of
specular tunneling \cite{Takayama} 
\begin{eqnarray}
\sum_k  \rightarrow  {\cal{A}} N_C(0) \int d \epsilon_k \\
\sum_{p \perp}  \rightarrow \rho_A(0) d_A \int d \epsilon_p
\label{approx}
\end{eqnarray}
where $\cal A$ is the area of the junction, $d_A$ is a width of
layer {\bf A},
$N_C(0)$ is a two dimensional 
density of states in layer {\bf C} and $\rho_A(0)$ is a one dimensional
density of electrons in layer {\bf A}.
Equations (\ref{explicit}) - (\ref{approx}) serve as the starting
point for our discussion of the tunneling measurement 
of the $\pi$ resonance.

Being a collective mode, the $\pi$ resonance is represented by a pole
in the four-leg vertex.  Its first contribution comes in
the third order of perturbation theory \footnote{From now on we will
  use a finite-temperature Matsubara technique}:
\begin{eqnarray}
\label{crnt-general}
J^{(3)}(\tau_0) & =& - \frac{1}{6}\int_{0}^{1/T} d\tau_1
        d\tau_2 d\tau_3 
        \Big\langle
           T_{\tau}\{ H_T(\tau_1) H_T(\tau_2) H_T(\tau_3) J(\tau_0)\} 
        \Big\rangle
\end{eqnarray}
with 
\begin{equation}
\label{crnt-def}
J(\tau)=- e  \sum_{kp,\alpha\beta} \left( T_{kp}^{\alpha\beta} 
a^{\dagger}_{k\alpha}(\tau) c_{p\beta}(\tau) e^{i \Omega \tau} -
T_{pk}^{\beta\alpha*}  
c^{\dagger}_{p\beta}(\tau) a_{k\alpha}(\tau) e^{-i \Omega \tau}
\right)_{i \Omega  \rightarrow eV } 
\end{equation}
Expression (\ref{crnt-general}) has many terms, each containing four
$a$ and four $c$ operators.  
The resonant contribution describe the coupling of the BCS condensate
in {\bf A} to the $\pi$ resonance in {\bf C}, and
is given by the terms which have
anomalous two-particle Green's function $K_L$ on the superconducting
side, and the two-particle Green's function $K_R$ on the normal-metal
side  that takes into 
account the multiple scattering of the particles of each other. The
anomalous $K_L$ has total momentum zero and total spin zero and the
singular part of $K_R$, as shown in \cite{neutron}, corresponds to the two
particles having the 
center of mass momentum $Q$ and spin one. This mismatch is compensated
by the matrix elements $T^f$ which flip the 
spin and add momentum $Q$ to the pair. 
Disposing all momentum, spin and energy conservation properties of the
bulk Green's function we come to the $\pi$ resonance contribution to the
tunneling current:
\begin{eqnarray}
J_{\pi} &=&  - \frac{4 e}{N} Im \{~  [ \sum_{kpk''p''}  T^{d}_{kp}
  T^{d*}_{k''p''} 
T^{f}_{-p-k}T^{f*}_{-p''-k''} ~\Gamma_{\uparrow}(k,k'';2 i \Omega,Q) \nonumber
\\ && T \sum_{ \omega_1} 
F_p(i \omega_1) G_k(-i \omega_1-i \Omega) G_{Q-k}(i \omega_1-i \Omega) \nonumber
\\&& T \sum_{\omega_2}
F_{p''}(i \omega_2) G_{k''}(-i \omega_2- \Omega) G_{Q-k''}(i \omega_2-
\Omega)~ ]_{i \Omega \rightarrow eV} \}
\label{j-gen}
\end{eqnarray}
where $\Gamma_{\uparrow}(k,k',E,Q)$ is the vertex for all spins up
with total energy $E$ and momentum $Q$.
$N={\cal A}/a^2$ 
with $a$ being a unit cell size of $YBCO$ . Similar expression
has been studied in \cite{Takayama} in connection with the problem of
the fluctuational contribution to the tunneling currents in
conventional superconductors. Diagrammatically expression (\ref{j-gen})
is shown on Figure 2.

\begin{figure}[h]
\centerline{\epsfysize=8cm
\epsfbox{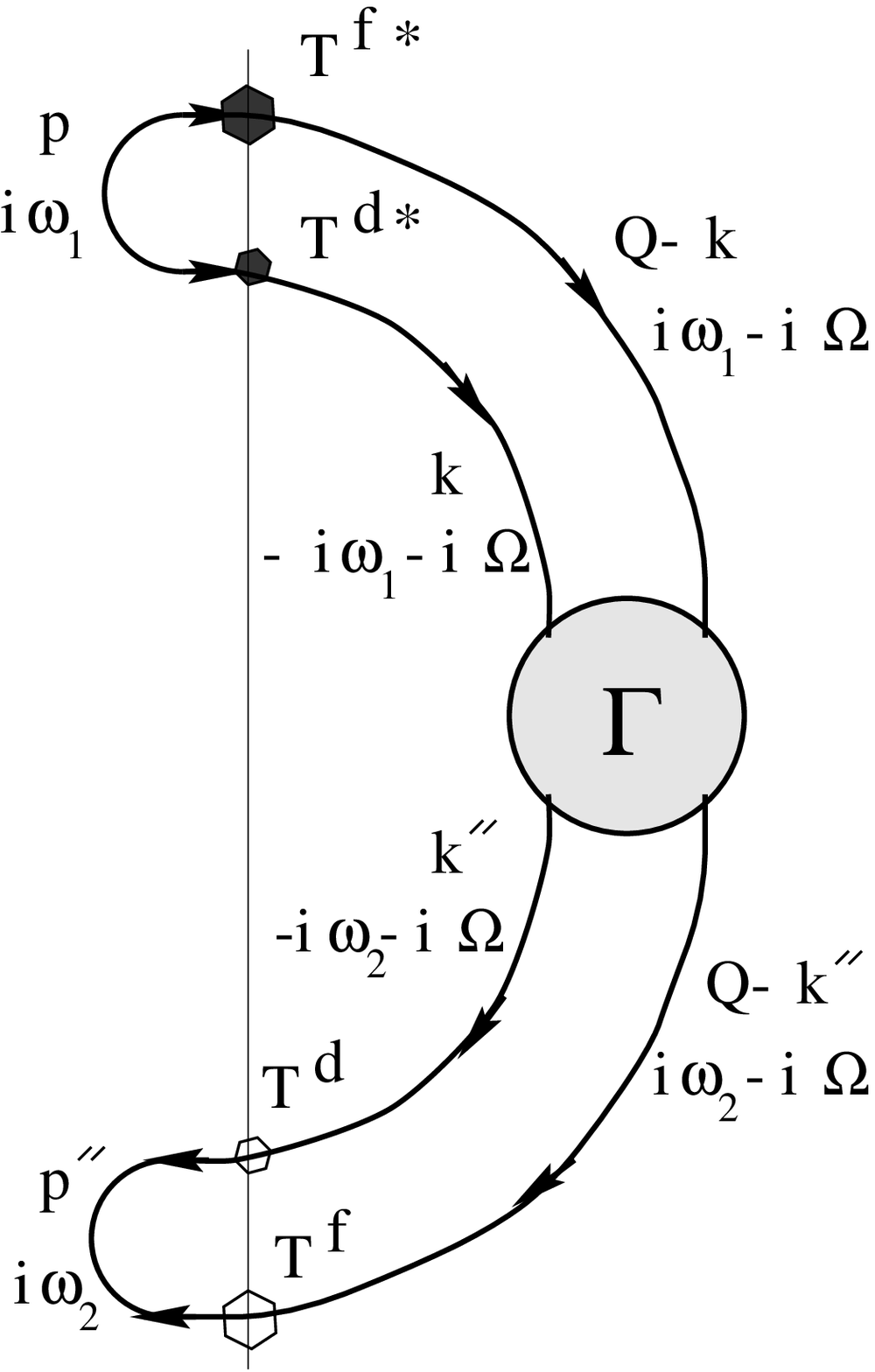}
}
\caption{Second order tunneling diagram.}
\end{figure}

We assume that the normal metal is described by the $t$-$J$
Hamiltonian and the triplet vertex $\Gamma_{\uparrow}$ may be found
from the Dyson's equation
\begin{eqnarray}
\Gamma_{\uparrow}( k;k''; 2 i \Omega, Q)= J \sum_{\alpha }
 g_{\alpha}(k)  g_{\alpha}(k'') 
 -J T  \sum_{k'\nu\alpha} g_{\alpha}(k) g_{\alpha}(k') G_{k'}(i \nu)
G_{Q-k'}(2 i \Omega- i \nu)  
\Gamma_{\uparrow} (k',k'';2 i \Omega,Q)
\label{Gamma-eq}  
\end{eqnarray}
where $\alpha$ is an index that may be $+$ or $-$ and $g_{\pm}(k) =
cos k_x \pm  cos k_y$.

The $t$-$J$ Hamiltonian possesses two remarkable 
properties. Namely,  the two particle continuum collapses to a point
when their center of mass momentum is
$Q$, and second there is repulsive interaction between the two particles
in a triplet state sitting on the neighboring sites. This leads to
the existence of an
antibonding state of two electrons with center of mass momentum $Q$
and energy $\omega_0 = J(1-n)/2-2\mu$ - the $\pi$ resonance\cite{neutron}.
Identifying this energy with the observed resonant neutron scattering 
peaks give $\omega_0=41meV$ or $33meV$ depending on the $T_c$ of layer
{\bf C}.
This antibonding state appears up as a sharp pole in 
$\Gamma_{\uparrow}(k,k', 2 i \Omega,Q)$  when $ 2 i \Omega = \omega_0
$. So, we can write solution to (\ref{Gamma-eq}) as
\begin{eqnarray}
\Gamma_{\uparrow}( k;k''; 2 i \Omega, Q) = J  g_{-}(k) g_{-}(k'')
\frac{2 i \Omega + 2 \mu}{2 i \Omega - \omega_0  } + \Gamma_{\uparrow}^{reg}
\end{eqnarray}

Putting together equations (\ref{j-gen}) and (\ref{Gamma-eq}) and
noticing that the anisotropic gap of the superconductor may be written
as $\Delta_p = \Delta~ g_{-}(p_{||})$ we get 
\begin{eqnarray}
J_{\pi}(V) &=& 2 e J \Delta^2 Im\{ ~[~\frac{ 2 i \Omega+2 \mu}{ 2 i \Omega-
  \omega_0  }  
~\frac{1}{N} \sum_{kpk''p''} T^{d}_{kp} T^{d*}_{k''p''}
T^{f}_{-p-k}T^{f*}_{-p''-k''} \nonumber 
\\ &&\hspace{2cm} g_{-}(k'') g_{-}(p_{||}'') g_{-}(k) g_{-}(p_{||}) 
R( i\Omega,E_p,\epsilon_k) R(i
  \Omega,E_{p''},\epsilon_{k''}) ~]_{i \Omega 
  \rightarrow 
  eV} \} 
\label{j_pi_1}
\end{eqnarray}
where
\begin{eqnarray}
R(i \Omega,E_p,\epsilon_k) &=&
\frac{ tanh( E_p/T ) }{ E_p ( E_p - \epsilon_k - i \Omega ) ( E_p - 2
  \mu - \epsilon_k + i \Omega )} \nonumber\\
&+& \frac{ tanh( \epsilon_k/T ) }{( E_p - \epsilon_k - i \Omega ) ( E_p -
  2 \mu - \epsilon_k + i \Omega ) ( 2 \mu - 2 i \Omega )}
\nonumber\\
E_p &=& \sqrt{ \epsilon_p^2 + \Delta_p^2} 
\end{eqnarray}
In the expression above the angular dependence comes from $g_{-}$
functions as well as from the anisotropy of $\Delta_p$ in the
expressions for $E_p$. It is easy to convince oneself that the later
does not have any significant effect on the result. So, we can neglect
the anisotropy of $\Delta_p^2$ and replace it by the average value
$\Delta^2$. Then integrating over directions of $k$ and $k''$ may be
done explicitly giving the average values of $ \langle g^2_{-}(k) \rangle \approx
1$. Finally we arrive at the following expression for $J_{\pi}$
\begin{eqnarray}
J_{\pi} &=& \frac{2 e J \Delta^2}{N} (T^d T^f~ A~ d_A~
 \rho_A(0)~ N_C(0)~ )^2  
 Im\{ \frac{ 2 e V +2 \mu}{ 2 e V - \omega_0 + i 0 } 
\left ( \int d \epsilon_p d \epsilon_k R( eV + i 0,E_p,\epsilon_k)
  \right )^2  \} 
\label{j_pi_2}
\end{eqnarray}

On Fig.3 we present the characteristic $V$ dependence of
$j_{\pi}$. One can see that it does have a sharp pole when
$eV_0=\omega_0/2$ which, if found, will be a clear indication of the
existence of the $\pi$ excitation in $YBCO$ materials. 
\begin{figure}[h]
\centerline{\epsfysize=6cm
\epsfbox{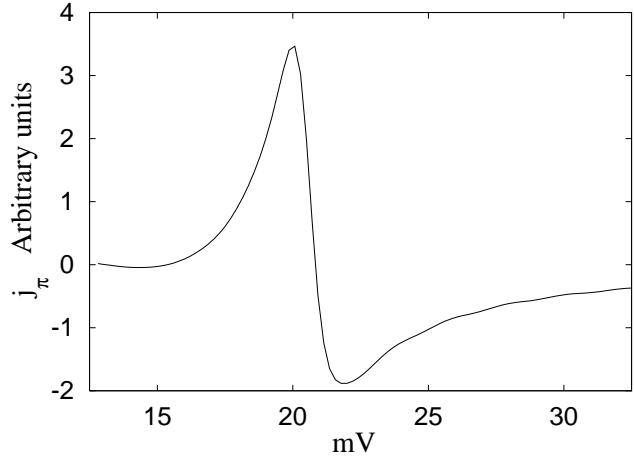}
}
\caption{$j_{\pi}$ as a function of V.}
\end{figure}

We can do a simple estimate of the integrated
spectral weight of the $\pi$ resonance. The usual expression for the
normal-to-normal tunneling current is given by 
\begin{eqnarray}
J_N & = &  e \sum_{pk} | T^{d}_{pk}|^2 \delta( eV + \epsilon_k -\epsilon_p )
[ n_F (\epsilon_k ) - n_F (\epsilon_p ) ] \nonumber\\
& = &  d_A {\cal A} | T^{d} |^2 \rho_C(0) N_A(0) e^2 V
\label{j_n}
\end{eqnarray}
Then, assuming that the characteristic scale of $\epsilon_p$'s and
$\epsilon_k$'s in 
equation (\ref{j_pi_2}) is set by $J$
we can obtain 
after a few straightforward manipulations 
\begin{eqnarray}
 \int J_{\pi} dV  \cong   \left| \frac{T^f}{T^d} \right|^2 
\frac {  \hbar a^2 {\cal A} \Delta^2 }{ e^4} 
\left( \frac{1}{R_N {\cal A}} \right)^2 
\end{eqnarray}
To get an idea of the magnitude of this effect we take the numbers
characteristic to the experiments of Goldman {\it et al. } on the
fluctuational contribution to the S-N current in low temperature
superconductors\cite{Goldman}. 
${\cal A} \approx 10^{-4} cm^2$, $R_N \approx 10^{-1} \Omega$ and 
characteristic to $YBCO$ gap
$\Delta=20 meV$ and $a=4.8 \times 10^{-8} cm$. For $T^f/T^d
\approx 1 $ this 
gives us $$ \int J_{\pi} dV  \approx 10~ \mu A~ \mu V $$
which is the effect of the same order of magnitude as 
measured by Goldman and coworkers\cite{Goldman}.

In conclusion we have proposed a concrete two particle tunneling
experiment to probe the $\pi$ resonance of the high $T_c$ superconductors.
Identification of this mode could uniquely distinguish among the various
theoretical explanations of the resonant neutron scattering peaks,
lend direct experimental support of the $SO(5)$ theory and deepen our
understanding of the symmetry relationship between antiferromagnetic and
superconducting phases in the high $T_c$ superconductors.  

We would like thank Prof. J. R. Schrieffer for suggesting the
possibility of detecting the pi resonance in the pair tunneling
process, and Prof. D. Scalapino for calling our attention to
references \cite{Scalapino} and \cite{Goldman}. This work is supported
by the National Science Foundation under grant numbers
DMR-9400372 and DMR-9522915.

\end{document}